\documentclass[
               12pt,
               aps,
               prd,
               amsmath,
               amssymb,
               longbibliography,
               nofootinbib,
               showpacs,
               superscriptaddress,
               color
               ]{revtex4-1}

\usepackage{graphicx}
\usepackage{hyperref}
\usepackage{lineno}
\usepackage{xspace}

\graphicspath{{figures/}}

\newcommand{\sherpa}{\textsc{Sherpa}\xspace}
\newcommand{\kT}{\ensuremath{k_{\mathrm{T}}}\xspace}
\newcommand{\pT}{\ensuremath{p_{\mathrm{T}}}\xspace}

\newcommand{\HF}{\ensuremath{\mathrm{HF}}\xspace}
\newcommand{\VHF}{\ensuremath{V+\mathrm{HF}}\xspace}
\newcommand{\ZHF}{\ensuremath{Z+\mathrm{HF}}\xspace}

\newcommand{\jet}{\ensuremath{\mathrm{jet}}\xspace}

\usepackage{color}

\begin{document}

\title{Hard production of a $Z$ boson plus heavy flavor jets
  at LHC and the intrinsic charm content of a proton}

\author{A.V.~Lipatov}
\affiliation{Skobeltsyn Institute of Nuclear Physics, Moscow State 
University, 119991 Moscow, Russia}
\affiliation{Joint Institute for Nuclear Research, 141980 Dubna, Moscow 
region, Russia}
\author{G.I.~Lykasov}
\affiliation{Joint Institute for Nuclear Research, 141980 Dubna, Moscow 
region, Russia}
\author{M.A.~Malyshev}
\affiliation{Skobeltsyn Institute of Nuclear Physics, Moscow State 
University, 119991 Moscow, Russia}
\author{A.A.~Prokhorov}
\affiliation{Joint Institute for Nuclear Research, 141980 Dubna, Moscow 
region, Russia}
\affiliation{Faculty of Physics, Lomonosov Moscow State University, 119991 
Moscow, Russia}
\author{S.M.~Turchikhin}
\affiliation{Joint Institute for Nuclear Research, 141980 Dubna, Moscow 
region, Russia}

\begin{abstract}
The cross section of associated production of
a $Z$ boson with heavy flavor jets in $pp$
collisions is calculated using the
\sherpa Monte Carlo generator and the analytical combined 
QCD approach based on \kT-factorization at small $x$ and
conventional 
collinear QCD at large $x$.
A satisfactory description of the ATLAS and CMS data on
the $\pT$ spectra of 
$Z$ bosons and $c$-jets in the whole rapidity, $y$, region is 
shown.
Searching for the intrinsic charm (IC) contribution in these processes, which 
could be visible at large $y> 1.5$, 
we study observables very sensitive to non-zero IC contributions and less 
affected by theoretical QCD scale uncertainties. 
One of such observables is the so-called
double ratio: the ratio of the  
differential cross section of $Z+c$ production in the central region of $|y|< 
1.5$ and in the forward region $1.5 < |y| < 2.5$, divided by the same ratio for 
$Z+b$ production.
These observables could be more promising for the search of IC 
at LHC as compared to the observables considered earlier. 

\end{abstract}

\pacs{12.15.Ji, 12.38.Bx, 13.85.Qk}

\maketitle

\section{Introduction}

Many hard processes within the Standard Model and
beyond, such as the production of heavy flavor jets, of the 
Higgs boson, and 
other processes are quite sensitive to the heavy
quark content of the nucleon.
Studying the latter plays an increasingly significant
role in the physics program of the Large Hadron Collider (LHC). 
Strange, charm and beauty parton distribution functions (PDFs)
are essential inputs for the calculation of observables for
these processes within 
the perturbative Quantum Chromodynamics (pQCD).   
 Global QCD
analysis allows one to extract the PDFs from comparison of hard-scattering data
 and pQCD calculations. 

Hard production of vector bosons accompanied by heavy 
flavor\footnote{Here and below heavy flavor implies charm and beauty 
quarks.} jets (\VHF)
in $pp$ collisions at LHC energies can be considered as an additional tool to 
study the quark and gluon PDFs compared to the deep inelastic scattering 
of electrons on protons.
In these processes, in the rapidity region $|y|<2.5$, which corresponds to 
the kinematics of ATLAS and CMS experiments,
one can study these PDFs not only at low parton momentum fractions 
$x<0.1$ but also at larger $x$ values~\cite{Brodsky:2016fyh}. 
Therefore, such \VHF processes can give us new 
information on the PDFs at large $x>0.1$, 
where the non-trivial proton structure (for example, the
possible contribution of valence-like \textit{intrinsic} heavy
quark components) can be 
revealed\cite{Brodsky:1980pb,Brodsky:1981se,Harris:1995jx,Franz:2000ee}.

Intense studies of an intrinsic charm (IC) signal in the
production of 
vector ($Z$ and $W$) bosons or prompt photons 
$\gamma$ accompanied by heavy-flavor jets in $pp$ collisions at 
LHC energies were  made in~\cite{Bednyakov:2013zta, 
Beauchemin:2014rya,Lipatov:2016feu,Brodsky:2016fyh,Ball:2016neh}. 
It was shown that the contribution of IC    
to the proton PDFs can be visible in the transverse momentum spectra 
of $\gamma/Z/W$
or $c/b$-jets in the forward rapidity region of the ATLAS and
CMS kinematics, 
$1.5< |y|< 2.5$, at large $\pT > 100$~GeV. The shape of these 
\pT spectra depends significantly
on the IC probability in the proton $w_{IC}$, while in
the more central rapidity 
region $|y|< 1.5$ the IC signal may not be visible.

Up to now there is a long-standing debate about the $w_{IC}$ value 
\cite{Ball:2016neh,Brodsky:2015uwa,Jimenez-Delgado:2014zga,Hou:2017khm}  
(see also \cite{Brodsky:2016fyh} and references therein).
A first estimate of the intrinsic charm probability in the
proton was carried out in~\cite{Bednyakov:2017vck} utilizing recent ATLAS
data on the production of prompt photons accompanied by $c$-jets at
$\sqrt{s}= 8$~TeV~\cite{Aaboud:2017skj}.
An upper limit $w_{IC} < 2.74$\% ($3.77$\%) at $68$\% ($95$\%) C.L. 
was set~\cite{Bednyakov:2017vck}.
It is shown~\cite{Bednyakov:2017vck} that to extract the IC probability
from these ATLAS data
we have to eliminate a large theoretical uncertainty due to the QCD scale.
In this paper we focus on looking for observables in \ZHF production 
processes, which are sensitive to the IC contribution in the
proton PDF and are less dependent on the QCD scale.
In \cite{Lipatov:2016feu} it
is shown that such observables could be the ratio of
$\gamma/Z+c$ and $\gamma/Z+b$ production cross sections in  
the forward rapidity region $1.5 < |y^{Z}| < 2.5$.
Calculations~\cite{Lipatov:2016feu} were performed 
applying the MCFM~\cite{Campbell:2002tg} Monte Carlo (MC)
generator and \kT-factorization of QCD.  

In this paper we investigate \ZHF production processes at LHC  
energies within two approaches: the combined QCD approach, 
based on the   
\kT-factorization formalism~\cite{Levin:1991ry,Catani:1990eg,Collins:1991ty} in 
the small $x$ domain and on conventional 
(collinear) QCD factorization at large $x$, and the  
\sherpa MC event generator~\cite{Gleisberg:2008ta}. 
Recently the combined QCD approach was successfully applied 
to describe LHC data on associated $Z + b$
production at $\sqrt{s} = 7$~TeV~\cite{Baranov:2017tig}. 
The \sherpa MC generator, 
which includes initial and final state parton showering, is
supposed to 
provide a realistic description of multi-particle final states 
allowing for \HF jets from higher perturbative orders, such as gluon splitting 
into heavy quark pairs. \sherpa can also
model the full chain of hadronization and decays of unstable particles,
that should allow us a more accurate comparison to experimental measurements of
\HF jets than achieved in previous 
studies~\cite{Beauchemin:2014rya,Lipatov:2016feu}.
Validation of these approaches is performed using ATLAS and CMS 
data~\cite{Aad:2014dvb, Sirunyan:2017pob} on $Z$ boson production accompanied 
by charm and beauty jets for center-of-mass 
energies $\sqrt{s} = 7$ and 8~TeV.  
One of the goals of this work is to study the influence of 
intrinsic charm on 
various kinematical distributions in these processes 
and to investigate the effects of initial and final state
parton showers in the description of LHC data.
We also focus on 
finding new observables which are sensitive to the IC content of a proton
and which could help us to reduce the QCD scale uncertainties. 

In Section II we present two theoretical approaches
adopted in
our calculations. The results and discussion are presented in 
Section III, and Section IV is the Conclusion.

\section{Theoretical approaches to associated \ZHF production}

To calculate the total and differential cross sections of associated \ZHF 
production within the combined QCD approach, we strictly
follow the 
scheme described earlier~\cite{Baranov:2017tig}. In this scheme, the 
leading contribution comes from the ${\cal O}(\alpha \alpha_s^2)$
off-shell gluon-gluon fusion subprocess $g^* + g^* \to Z + Q + \bar 
Q$ (where $Q$ denotes the heavy 
quark), calculated in the $k_T$-factorization approach.
The latter has certain technical advantages in 
the ease of including higher-order radiative corrections in the
form of transverse momentum dependent (TMD) parton 
distributions (see~\cite{Andersson:2002cf, Andersen:2003xj, Andersen:2006pg} 
for more information). To extend the consideration 
to the whole kinematic range, several subprocesses involving 
initial state quarks, namely flavor excitation 
$q + Q\to Z + Q + q$, quark-antiquark annihilation $q + \bar{q}\to Z + Q + 
\bar Q$ and quark-gluon scattering $q + g\to Z + q + Q \bar{Q}$, are taken into 
account using the collinear QCD factorization (in the tree-level 
approximation). 
The IC contribution is estimated using the 
${\cal O}(\alpha \alpha_s)$ QCD Compton scattering $c + g^* \to Z + c$,
where the gluons are kept off-shell but the incoming non-perturbative 
intrinsic charm quarks are treated 
as on-shell ones\footnote{The perturbative charm contribution is already taken 
into account in the off-shell gluon-gluon fusion subprocess.}.
Thus we rely on a combination of two 
techniques, with each of them being used for the kinematics
where it is more 
suitable\footnote{An essential point of 
consideration~\cite{Baranov:2017tig} is using a numerical solution of the CCFM 
evolution equation~\cite{Ciafaloni:1987ur,Catani:1989yc,Catani:1989sg} to
derive the  TMD  gluon density in a proton. The 
latter smoothly interpolates between the small-$x$ BFKL gluon dynamics and 
high-$x$ DGLAP dynamics. Following~\cite{Baranov:2017tig}, below 
we adopt the latest JH'2013 parametrization~\cite{Hautmann:2013tba}, 
adopting the JH’2013 set 2 gluon as the default
choice.}
(off-shell gluon-gluon fusion subprocesses at small $x$
and quark-induced subprocesses at large $x$ values).
More details of the above calculations can be found 
in~\cite{Baranov:2017tig}.

In contrast to earlier studies~\cite{Beauchemin:2014rya,Lipatov:2016feu} of \ZHF 
production within the MCFM routine (that
performs calculation in the fixed order 
of pQCD), in the present paper the
\sherpa 2.2.1~\cite{Gleisberg:2008ta} MC 
generator is applied. It uses matrix elements that are
provided by the built-in generators Amegic++~\cite{Krauss:2001iv} and 
COMIX~\cite{Gleisberg:2008fv}; 
OPENLOOPS~\cite{Cascioli:2011va} is used to introduce addtional loop 
contributions into the NLO calculations.
We use matrix elements calculated at the next-to-leading order (NLO) 
for up to 2 final partons and at the leading-order (LO) for up to 4 partons. 
They 
are merged with the \sherpa parton showering~\cite{Schumann:2007mg} following 
the 
ME+PS@NLO prescription~\cite{Hoeche:2012yf}. This is different from the study of 
$Z+c$ production carried out in~\cite{Hou:2017khm} where the matrix element was 
calculated in the LO and merged following the ME+PS@LO 
method~\cite{Hoeche:2009rj}. The 
latter approach was also used in this study as a cross-check, with the LO 
matrix 
element allowing for up to 4 final partons.
In both approaches, the five-flavor scheme (5FS) is used where $c$ and $b$ 
quarks 
are considered as massless particles in the matrix element and massive in both 
the initial and final state parton showers. 
\sherpa can also model the full
chain of hadronization and unstable particle decays for an accurate comparison 
with experimental measurements of
\HF jets.

\section{Results and discussion} 
\subsection{Comparison with the LHC data at $\sqrt s = 7$ and $8$~TeV}
In this section we present comparisons of our calculations for \ZHF 
production made with the \sherpa generator and within the
combined QCD approach to the LHC Run~1 data,
in order to verify the applicability of these 
approaches for further predictions. Following~\cite{Bednyakov:2013zta, 
  Beauchemin:2014rya,Lipatov:2016feu,Brodsky:2016fyh,Ball:2016neh}, we
mainly concentrate on the transverse momentum
distributions of $Z$ bosons and/or \HF jets, where the IC effects are 
expected to appear\footnote{Recent ATLAS and CMS experimental data
  on associated $Z + b$ production taken at $\sqrt{s} = 7$~TeV
  as functions of other kinematical variables within the
  framework of the combined QCD approach are considered
  in~\cite{Baranov:2017tig}.}.

The first comparison is performed for associated $Z + b$
production measured by the ATLAS Collaboration~\cite{Aad:2014dvb} at
$\sqrt{s}=7$~TeV. According to~\cite{Aad:2014dvb}, the
following selection criteria were applied to generated events. Two leptons
originating from the $Z$ boson decay are 
required to have an invariant mass 
$76~\mathrm{GeV} < m_{\ell\ell} <106$~GeV with a minimum transverse momentum of 
each lepton $\pT^\ell > 20$~GeV and rapidity $|y^{\ell}|<2.5$.
In \sherpa generated events, jets are built using all stable particles
excluding 
the lepton pair from the $Z$ boson decay with the 
anti-\kT algorithm with a size parameter $R = 0.4$. They are required to have 
a rapidity $|y^{\jet}| < 2.4$ 
and minimum transverse momentum $\pT^\jet > 20$~GeV. Each jet is also required 
to be separated from any of the two leptons by $\Delta{R}_{\jet,\ell} > 0.5$. 
Jets are identified as $b$-jets, if there is a weakly decaying 
$b$ hadron with a transverse momentum $\pT^b > 5$~GeV within a cone $\Delta{R} 
= 0.3$ around the jet direction.
The same kinematic requirements are applied to final state
$b$ quarks 
(treated as $b$-jets at a parton level) when using the
combined QCD approach. \sherpa results were obtained 
within the ME+PS@NLO model. In both approaches
the CTEQ66 PDF 
set~\cite{Nadolsky:2008zw} was used.

\begin{figure}
\begin{center}
{\includegraphics[width=.5\textwidth]{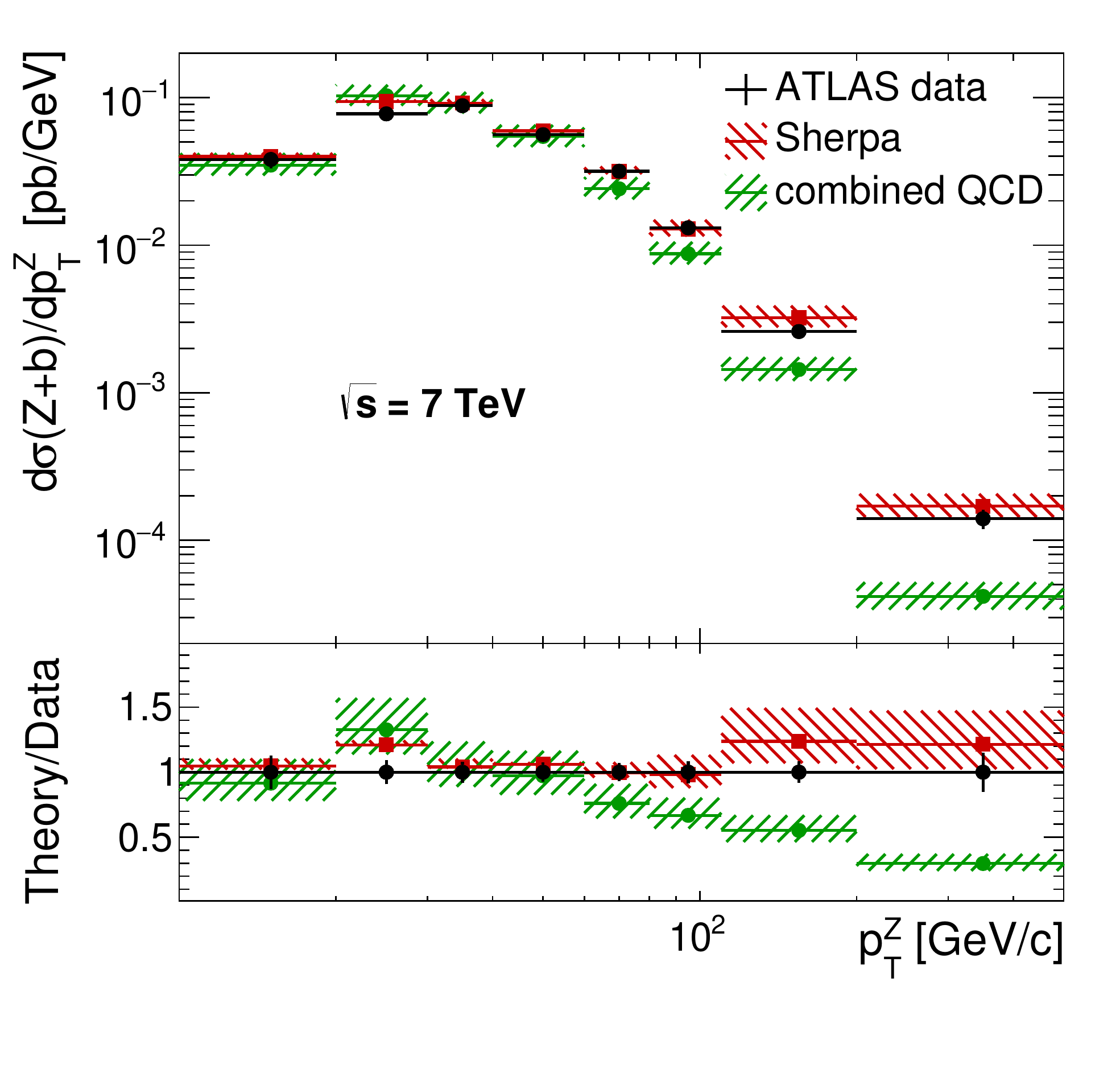}}
\caption{Cross section of $Z+b$-jet production as a function of
  the $Z$ boson 
transverse momentum in the full rapidity region $|y^{Z}| < 2.5$ at $\sqrt{s} = 
7$~TeV. The main panel shows the ATLAS measurement result~\cite{Aad:2014dvb} 
compared to \sherpa calculations and to combined QCD
calculations. The uncertainty bands represent
uncertainties in the QCD scale.
The bottom panel shows the ratio of calculations to data.}
\label{fig:Zb-7TeV}
\end{center}
\end{figure}

In Fig.~\ref{fig:Zb-7TeV} the associated $Z + b$-jet production cross 
section (for events with at least one $b$-jet) calculated as a function of
the $Z$ boson transverse momentum $\pT^Z$ is presented in
comparison with the ATLAS data~\cite{Aad:2014dvb}.
Here and below central values, marked by horizontal lines, correspond to 
the default choice of factorization and renormalization
scales $\mu_R = \mu_F = m_\mathrm{T}$, where $m_\mathrm{T}$ is the $Z$ boson
transverse mass. Theoretical uncertainties of our calculations correspond to
the maximum deviation between the nominal spectrum and those obtained by
the usual factor 2 variations of renormalization and
factorization scales. 

One can see that the \sherpa results are
in perfect agreement with the ATLAS data within the scale
uncertainties in the whole $\pT^Z$ range. In the combined QCD approach, we
observe some underestimation of the data at high $\pT^Z$ and a slight
overestimation at low transverse momenta. The latter can be attributed to the
TMD gluon density used in the calculations, because the region
$\pT^Z < 100$~GeV is fully dominated by the off-shell
gluon-gluon fusion subprocess~\cite{Baranov:2017tig}.
However, the results obtained within both approaches under consideration in 
this region are rather close to each other.
A noticeable deviation of the combined QCD calculations from the data at 
large $\pT^{Z}$ is explained by the absence of
the effects of parton showers, hadronization and additional contributions of 
NLO diagrams, including loop ones, in these calculations.
Such contributions, which are taken into account by \sherpa,
considerably improve the description of data.
The influence of the parton showers and of higher-order pQCD
corrections is investigated in detail in the next Section.
It is important to note that our results obtained with \sherpa are
in good 
agreement with the results obtained within a similar 
approach~\cite{Krauss:2016orf}.

Now, we turn to the associated $Z + c$-jet production
measured by the CMS Collaboration at $\sqrt{s} = 
8$~TeV~\cite{Sirunyan:2017pob}.
The following selection criteria are applied to generated events for this 
comparison. Two leptons originating from a $Z$ boson decay
must have an invariant mass 
$71~\mathrm{GeV} < m_{\ell\ell} <111$~GeV, a minimum transverse momentum of 
$\pT^\ell > 20$~GeV and rapidity $|y^{\ell}|<2.1$. Jets built with
the anti-\kT algorithm with a size parameter $R=0.5$ are
required to have $\pT^\jet > 
25$~GeV and $|y^{\jet}| < 2.5$ and to be separated from the leptons by 
$\Delta{R}_{\jet,\ell} > 0.5$. Similar $b$ and $c$ flavor identification 
criteria to those described above are used.

\begin{figure}
\begin{center}
{\includegraphics[width=.5\textwidth]{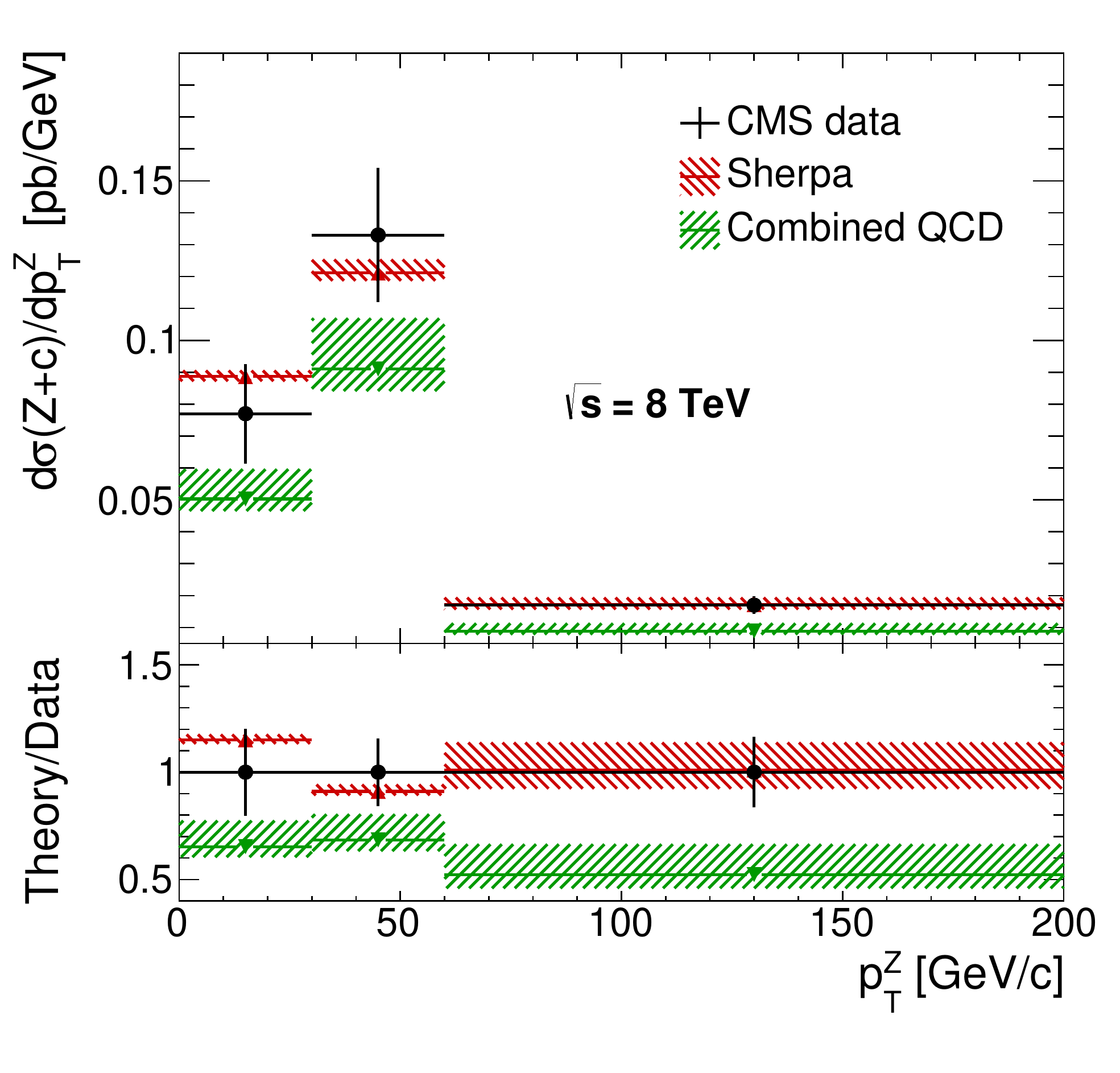}}\hfill
{\includegraphics[width=.5\textwidth]{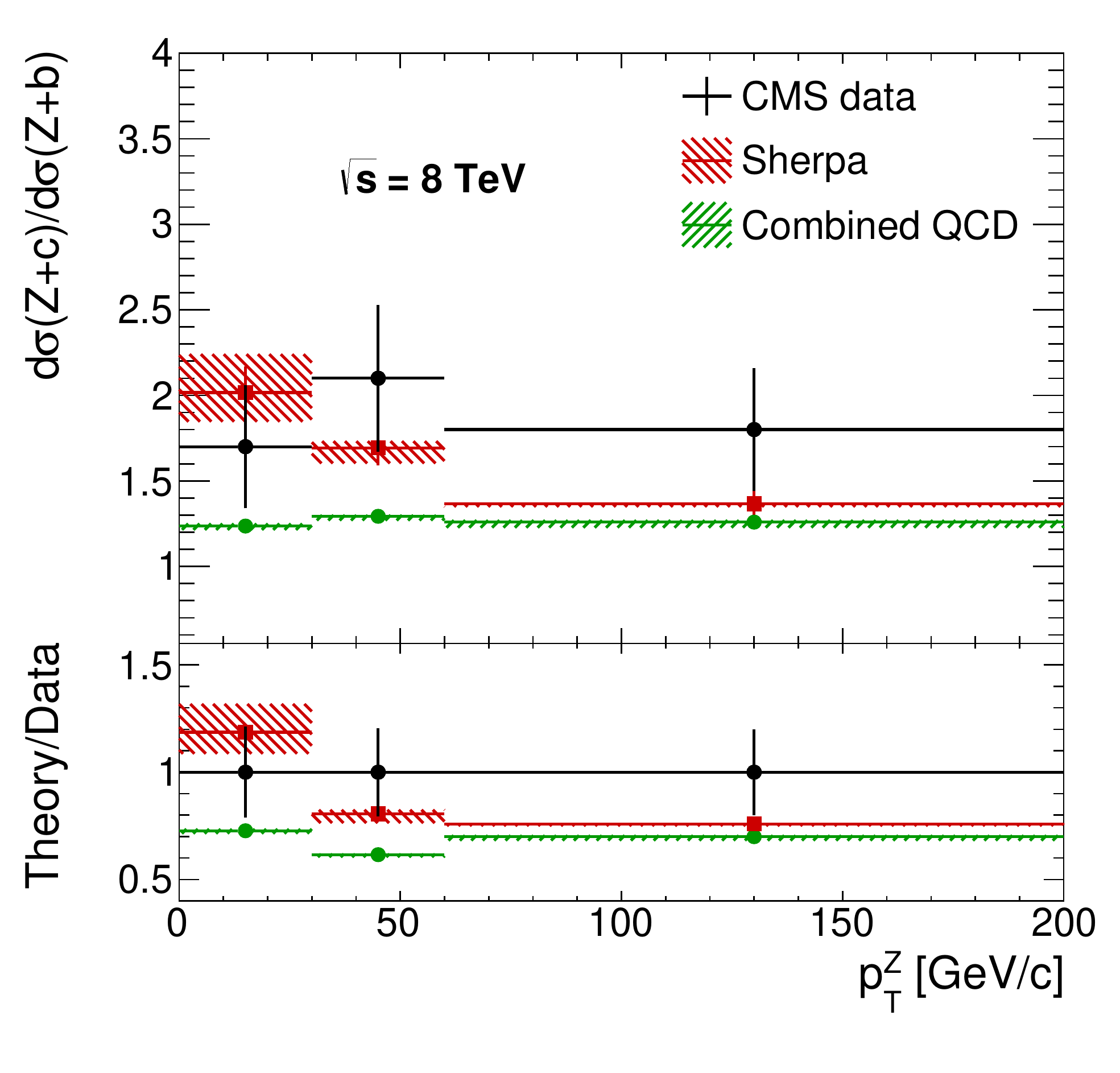}}\hfill\\
{\includegraphics[width=.5\textwidth]{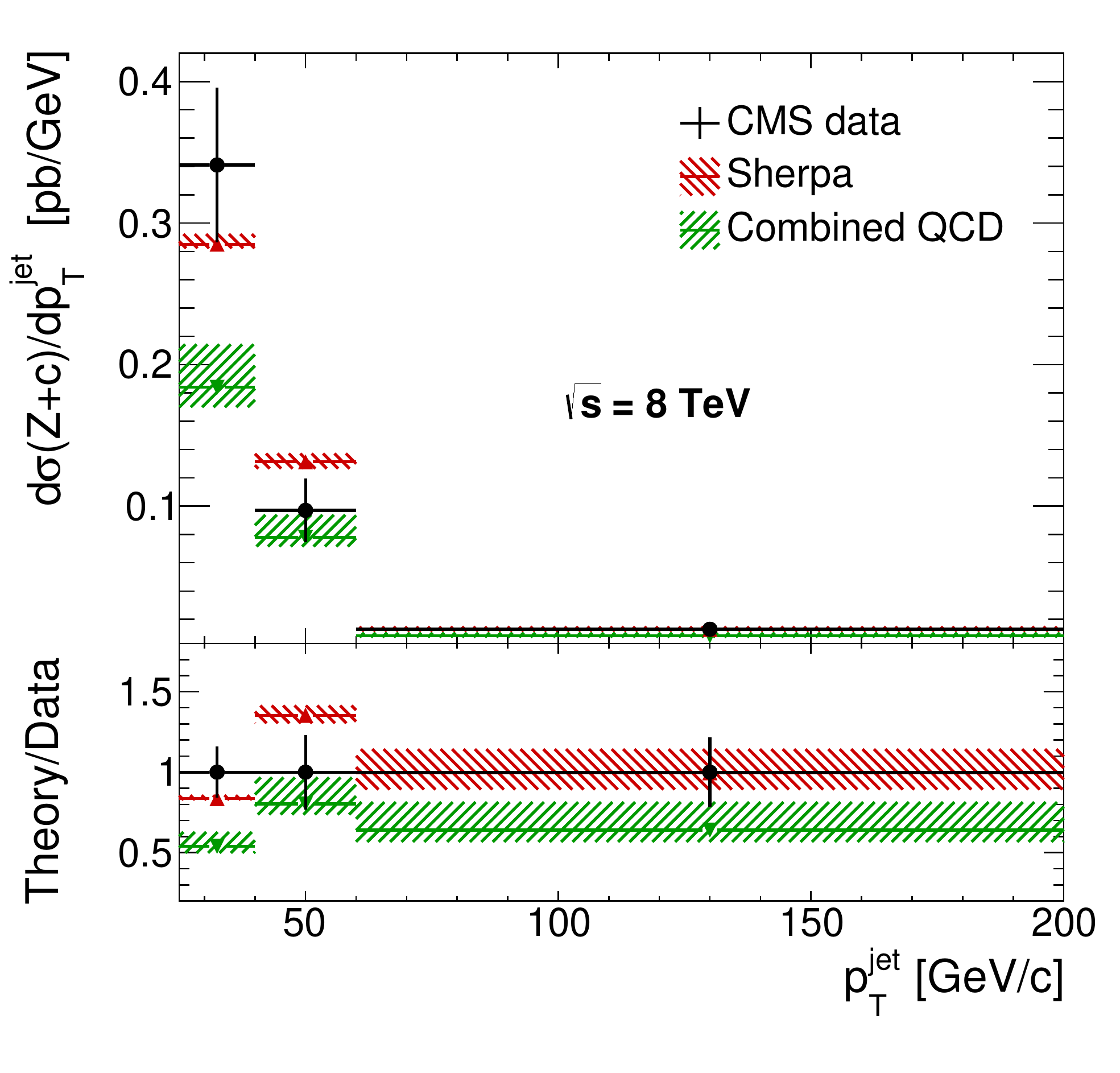}}\hfill
{\includegraphics[width=.5\textwidth]{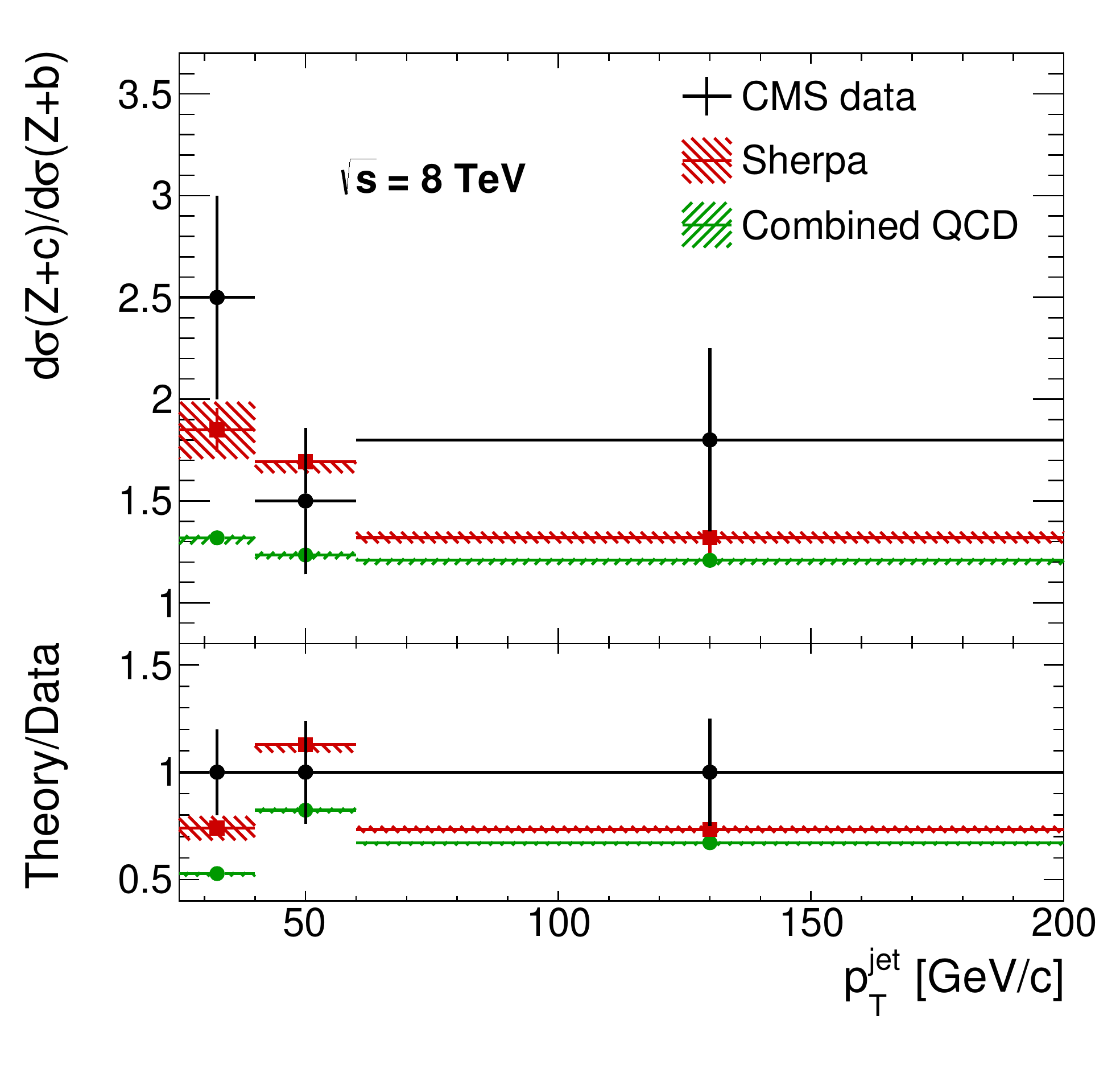}}\hfill
\caption{
Cross section of $Z+c$-jet production (left) and the ratio of cross sections 
of $Z+c$-jet and $Z+b$-jet production (right) as a function
of the $Z$ boson (top) and HF jet (bottom)
transverse momenta in the full rapidity region
$|y^{Z}| < 2.5$ at $\sqrt{s} = 8$~TeV. The main panels show the CMS
measurement result~\cite{Sirunyan:2017pob} compared to results of
\sherpa and the combined QCD calculations. The 
uncertainty bands represent the uncertainties in the QCD
scale. The bottom panels show the ratio of calculations to data.}
\label{fig:Zc-8TeV}
\end{center}
\end{figure}

In Fig.~\ref{fig:Zc-8TeV} our results for the differential cross 
sections of associated $Z + c$-jet production calculated as functions of the $Z$ boson and $c$-jet 
transverse momenta are shown in comparison 
with the CMS data~\cite{Sirunyan:2017pob}. 
A comparison with the measured ratio of the cross sections $\sigma{(Z + 
c)}/\sigma{(Z + b)}$ is also presented.
We find that the particle-level \sherpa calculations agree well with the data.
The parton-level combined QCD calculations also
describe the CMS data within the theoretical and experimental uncertainties 
(except at low $\pT^c < 40$~GeV),
although they tend to underestimate the \sherpa results.
As in the case of associated $Z + b$-jet production, we
attribute the latter to the parton showering effects and
additional NLO contributions, missing in the combined QCD calculations (to be
precise, mainly in the tree-level quark-induced subprocesses, since the
off-shell gluon-gluon fusion only gives a negligible
contribution at large transverse momenta). Note that the scale uncertainties
of our calculations partially cancel out when considering the
$\sigma{(Z + c)}/\sigma{(Z + b)}$ ratio\footnote{This ratio, being considered
in the forward rapidity region $1.5 < y^Z < 2.5$, is sensitive to the IC
content of a proton~\cite{Lipatov:2016feu}. However, we checked that in the
kinematical region probed by the CMS experiment~\cite{Sirunyan:2017pob} this
IC dependence is negligible.} (see Fig.~\ref{fig:Zc-8TeV}, right plots).

One can see that a better description of the CMS data is
achieved with the \sherpa 
tool and, therefore, we consider \sherpa calculations as the most reliable ones.
Thus, we mainly concentrate on them when investigating the
possible effects from IC in the LHC experiments below.

\subsection{$Z$+HF spectra for $\sqrt{s} = 13$~TeV and prediction
  for the IC contribution}
The purpose of the calculation of \ZHF differential cross
sections in this paper is to investigate the
effect of an IC signal on the observables, which can be
measured at the LHC by general purpose detectors at
$\sqrt{s} = 13$~TeV. As it was mentioned above, a sensitivity
to the IC at ATLAS and CMS experiments on $Z+c$-jet production
can be achieved in the forward rapidity region $1.5 < |y^{Z}| < 2.5$ and
$\pT^Z > 50$~GeV~\cite{Beauchemin:2014rya, Lipatov:2016feu}. In this
kinematical region the shape of the 
$\sigma{(Z + c)}/\sigma{(Z + b)}$ ratio is sensitive to  
effects of IC and is less affected by scale
uncertainties than those of the transverse momentum spectra.
This fact provides an opportunity to measure
the IC contribution.

In \sherpa, predictions for \ZHF production are calculated
within the ME+PS@NLO 
model using the CT14nnlo PDF set~\cite{Hou:2017khm}
containing PDFs with IC probabilities $w_{IC} = 0$,~$1$ and
$2$\%~\cite{Hou:2017khm}. 
The following selection criteria are used in this analysis.
Two leptons from the $Z$ boson decay are required to have
a mass  
$76~\mathrm{GeV} < m_{\ell\ell} <106$~GeV, transverse momentum 
$\pT^\ell > 28$~GeV and rapidity $|y^{\ell}|<2.5$. Jets 
are reconstructed from  all stable particles, excluding the  
leptons,  with the anti-\kT algorithm with parameters $R = 0.4$ and
are required to have $|y^{\jet}| < 2.5$ and
$\pT^{\jet} > 20$~GeV, $\Delta{R}_{\jet,\ell} > 0.4$. The
identification of heavy flavor jets is performed as
follows. If there is a weakly decaying $b$ hadron 
with $\pT^{b} >5$~GeV within a cone of $\Delta{R} = 0.5$ around the jet 
direction, the jet is identified as a $b$-jet. If it is not
identified as such, the same criteria are applied for $c$-hadrons, and the jet
is identified as a $c$-jet, if one is found.

\begin{figure}
\begin{center}
{\includegraphics[width=.5\textwidth]{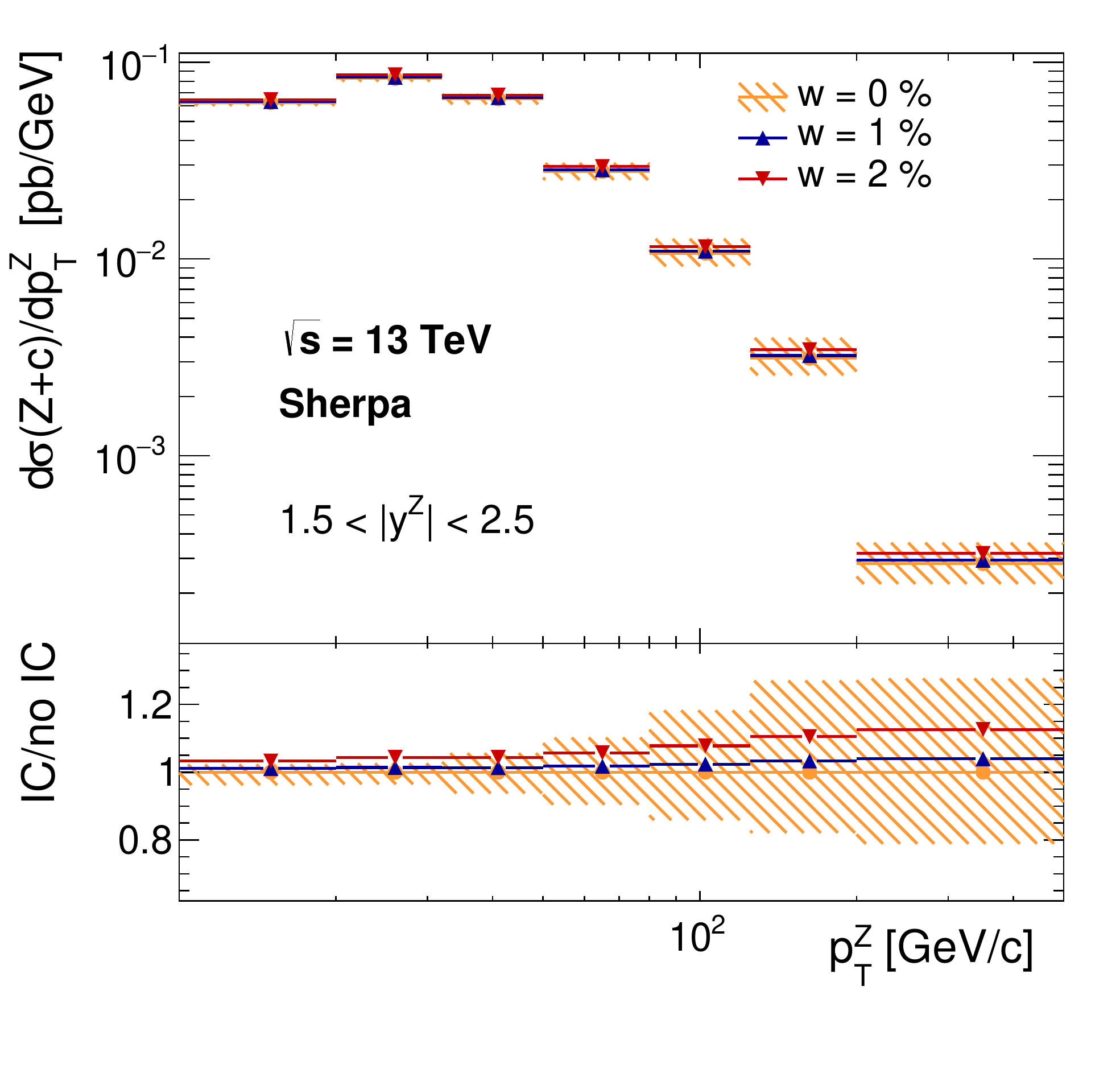}}\hfill
{\includegraphics[width=.5\textwidth]{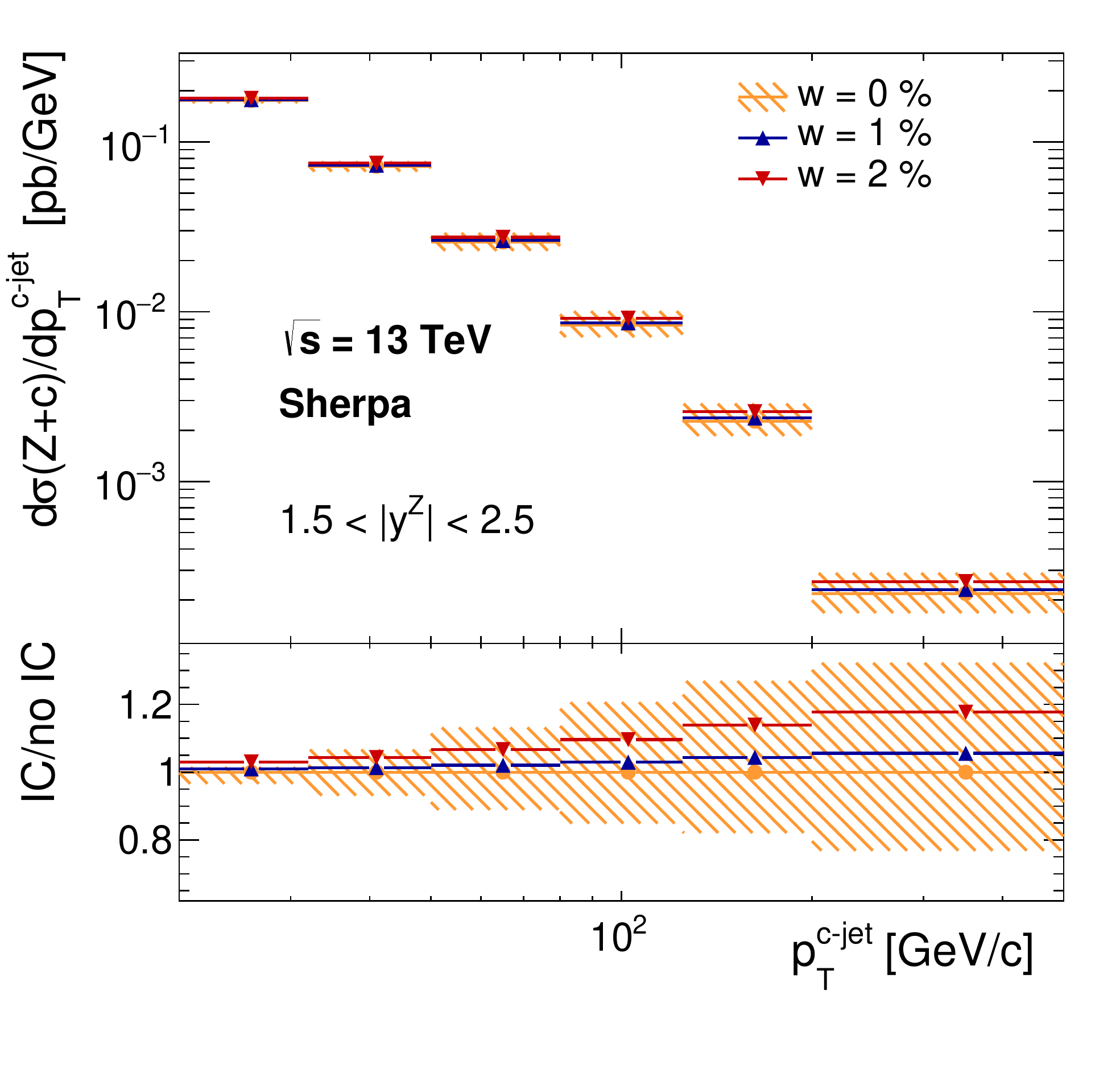}}\hfill
\caption{
Predictions for the cross section of $Z+c$-jet production as a function of
the $Z$ boson (left) and $c$-jet (right) transverse momentum
in the forward rapidity region 
$1.5 < |y^{Z}| < 2.5$ at $\sqrt{s} = 13$~TeV. The predictions are made with
the \sherpa generator using the CT14nnlo
PDF with different values for the IC contribution $w$. The
bottom panels show the ratio of predictions for non-zero values of $w$ 
to those for $w=0\%$. The uncertainty bands represent the uncertainties
in the  QCD scale (shown only for $w=0\%$ predictions). 
}
\label{fig:Zc_Sherpa_predictions}
 \end{center}
\end{figure}

\begin{figure}
\begin{center}
{\includegraphics[width=.5\textwidth]{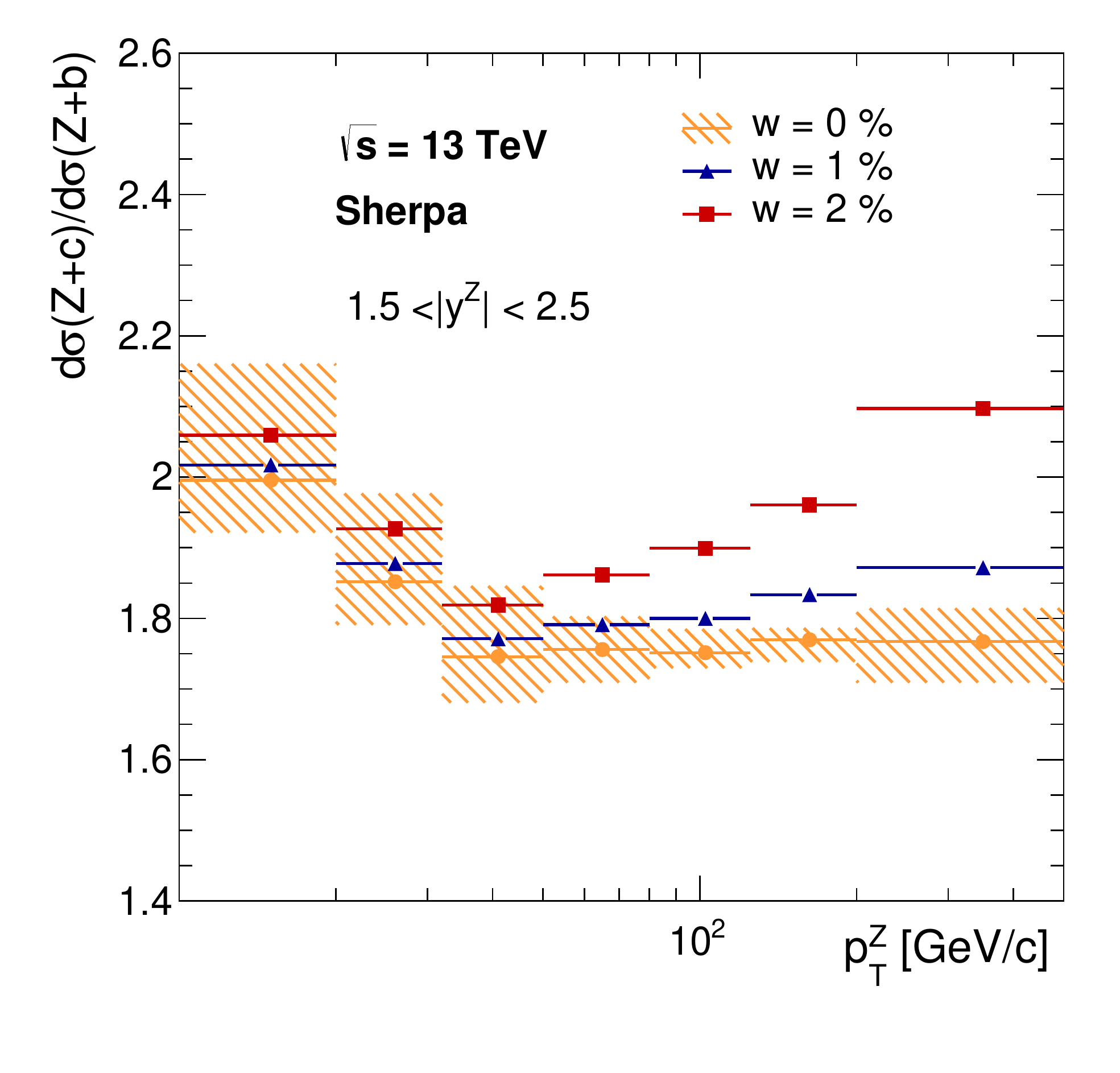}}\hfill
{\includegraphics[width=.5\textwidth]{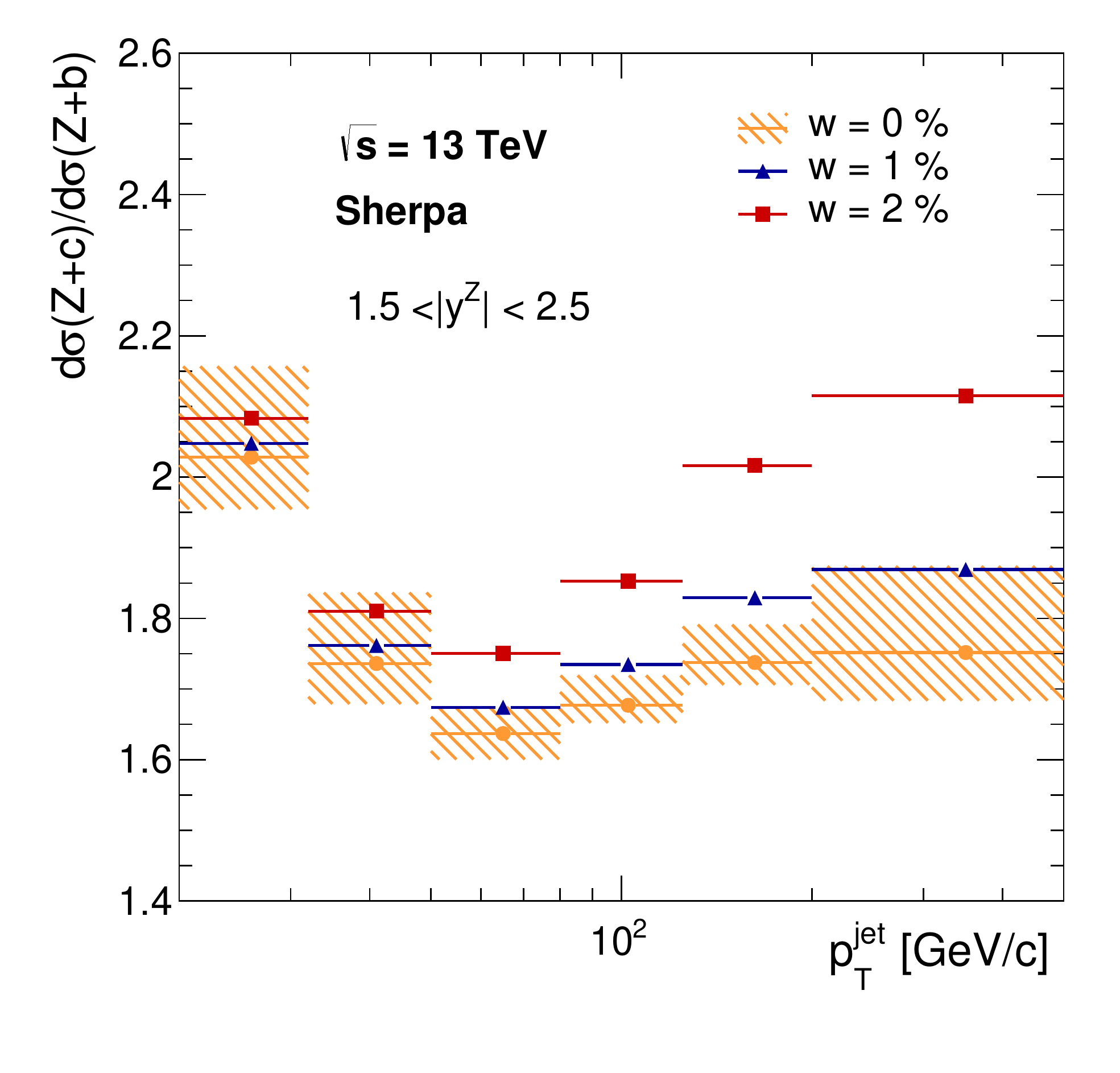}}\hfill
\caption{
Predictions for the ratio of $Z+c$-jet and $Z+b$-jet 
production cross sections as a function of
the $Z$ boson (left) and $c$-jet (right) transverse
momenta in the forward rapidity region  
$1.5 < |y^{Z}| < 2.5$ at $\sqrt{s} = 13$~TeV. The predictions are made with  
the \sherpa generator using the CT14nnlo PDF
with different values of the IC contribution $w$. The
uncertainty bands represent the uncertainties in the QCD
scale (shown only for $w=0\%$ predictions). 
}
\label{fig:Sherpa_ratios}
\end{center}
\end{figure}

In Fig.~\ref{fig:Zc_Sherpa_predictions} differential cross sections
of associated $Z + c$-jet production calculated in the 
forward rapidity region $1.5 < |y^{Z}| < 2.5$ at $\sqrt{s} = 13$~TeV as  
functions of the $c$-jet and $Z$ boson transverse momenta
are shown. The effect of IC becomes visible at $\pT \gtrsim 200$~GeV in both
distributions, but the theoretical uncertainties are still higher than the
size of this effect in the whole transverse momentum region 
studied. However, in the ratios of differential cross
sections $\sigma{(Z + c)}/\sigma{(Z + b)}$ the effect of IC can be visible at
significantly lower $Z$ boson or \HF jet
transverse momenta than in the differential cross sections
themselves. Predictions for these ratios are shown in
Fig.~\ref{fig:Sherpa_ratios}.

\begin{figure}
\begin{center}
{\includegraphics[width=.5\textwidth]{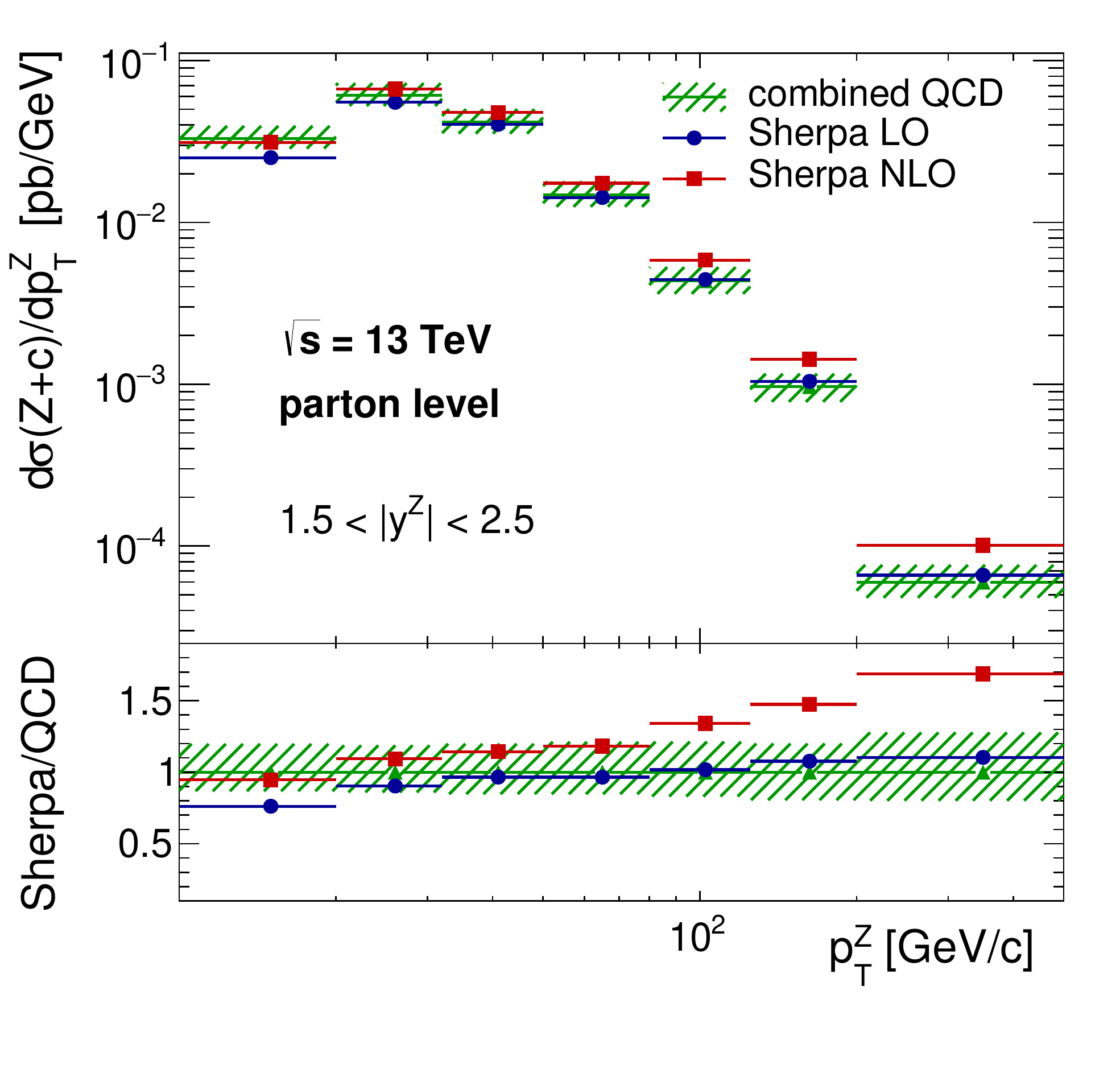}}\hfill
{\includegraphics[width=.5\textwidth]{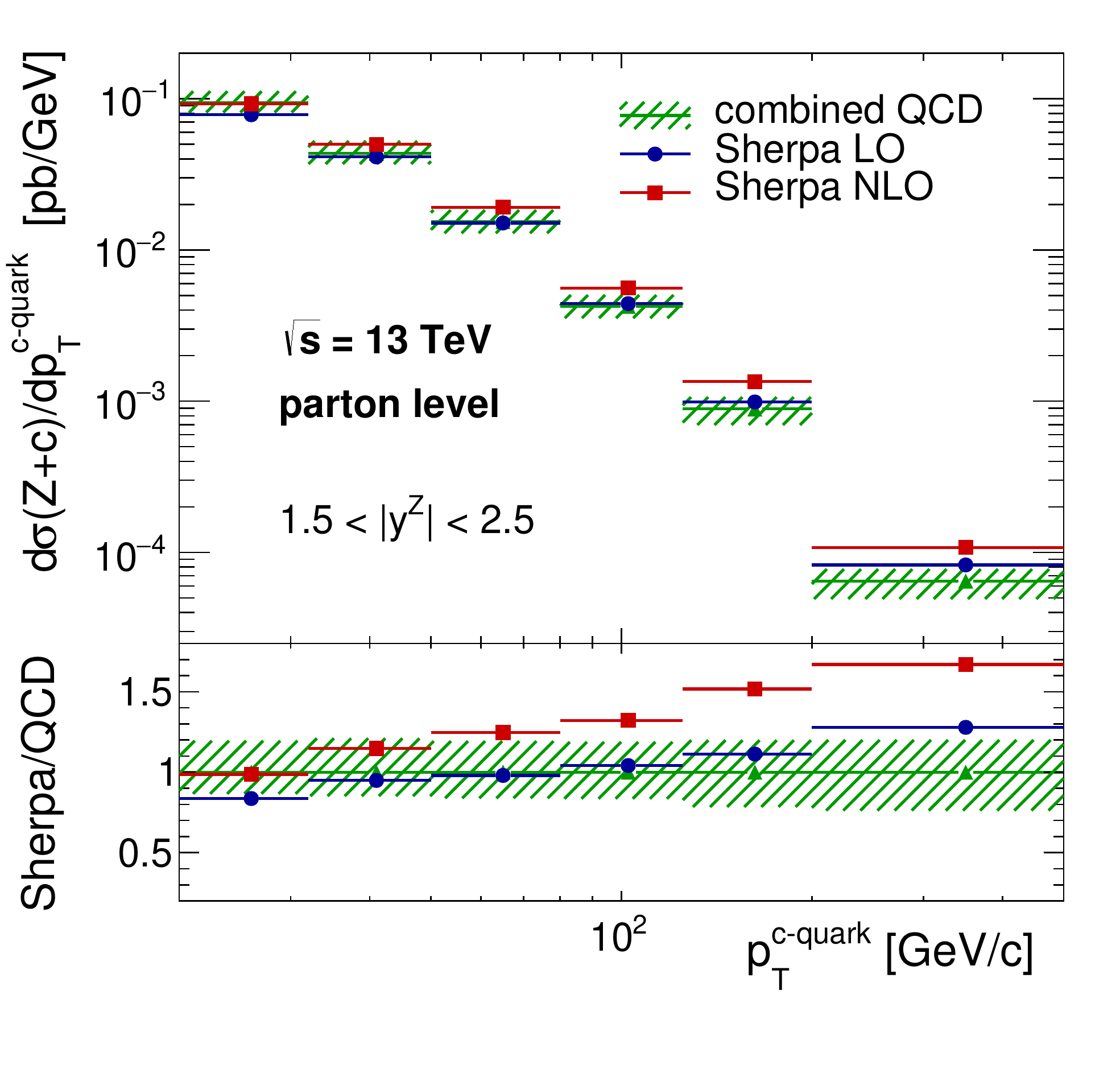}}\hfill
\caption{
Parton level predictions for the production cross section of
a $Z$ boson with a $c$ 
quark as a function of the $Z$ 
boson (left) and $c$-quark (right) transverse momenta in the
forward rapidity region 
$1.5 < |y^{Z}| < 2.5$ at $\sqrt{s} = 13$~TeV. The predictions are made
by combined QCD calculations and the  
\sherpa generator using LO and NLO matrix elements. The
CTEQ66 PDF set without any intrinsic charm contribution is
used. The uncertainty bands represent the uncertainties in the 
QCD scale (shown only for combined QCD predictions).
}
\label{fig:Comb_QCD_Sherpa_comp_13TeV}
\end{center}
\end{figure}

To investigate the influence of parton showers and higher-order 
pQCD corrections on the predictions, we repeated the above
\sherpa calculations at a parton level using LO and NLO
matrix elements. The results of these calculations are 
shown in Fig.~\ref{fig:Comb_QCD_Sherpa_comp_13TeV}
in comparison with the combined QCD predictions. First, one
can see that the 
best agreement with the combined QCD approach 
at large transverse momenta is given by the \sherpa calculations using the LO
matrix element. This is not surprising
because the combined QCD predictions are represented in this kinematical region
by the quark-induced subprocesses calculated in the usual collinear QCD
factorization with the same accuracy. At low and moderate transverse momenta 
the results of the combined QCD approach are consistently
close to parton-level \sherpa predictions obtained at the NLO level,
that demonstrates it is effective to take into account
higher-order pQCD corrections in the off-shell gluon-gluon fusion subprocess
supplemented with the CCFM gluon dynamics. Therefore, we can conclude that 
there are no large contradictions 
between our two theoretical approaches at the parton level. 
The combined QCD approach can be used to predict \ZHF
production cross sections at the  
parton level at moderate transverse momenta, but 
such approximation becomes worse towards high transverse momenta where the 
effects described above are quite large.

\begin{figure}
\begin{center}
{\includegraphics[width=.5\textwidth]{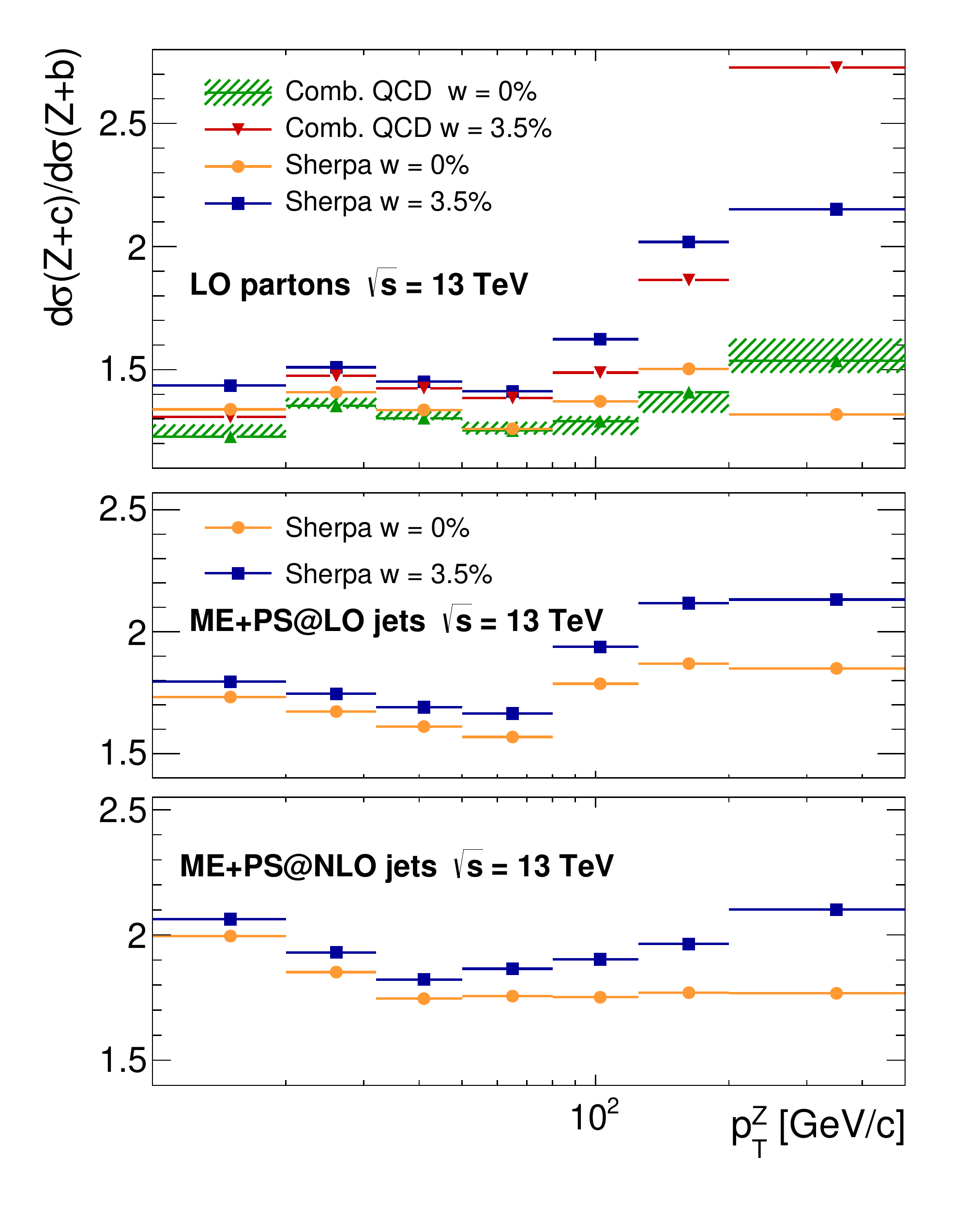}}\hfill
{\includegraphics[width=.5\textwidth]{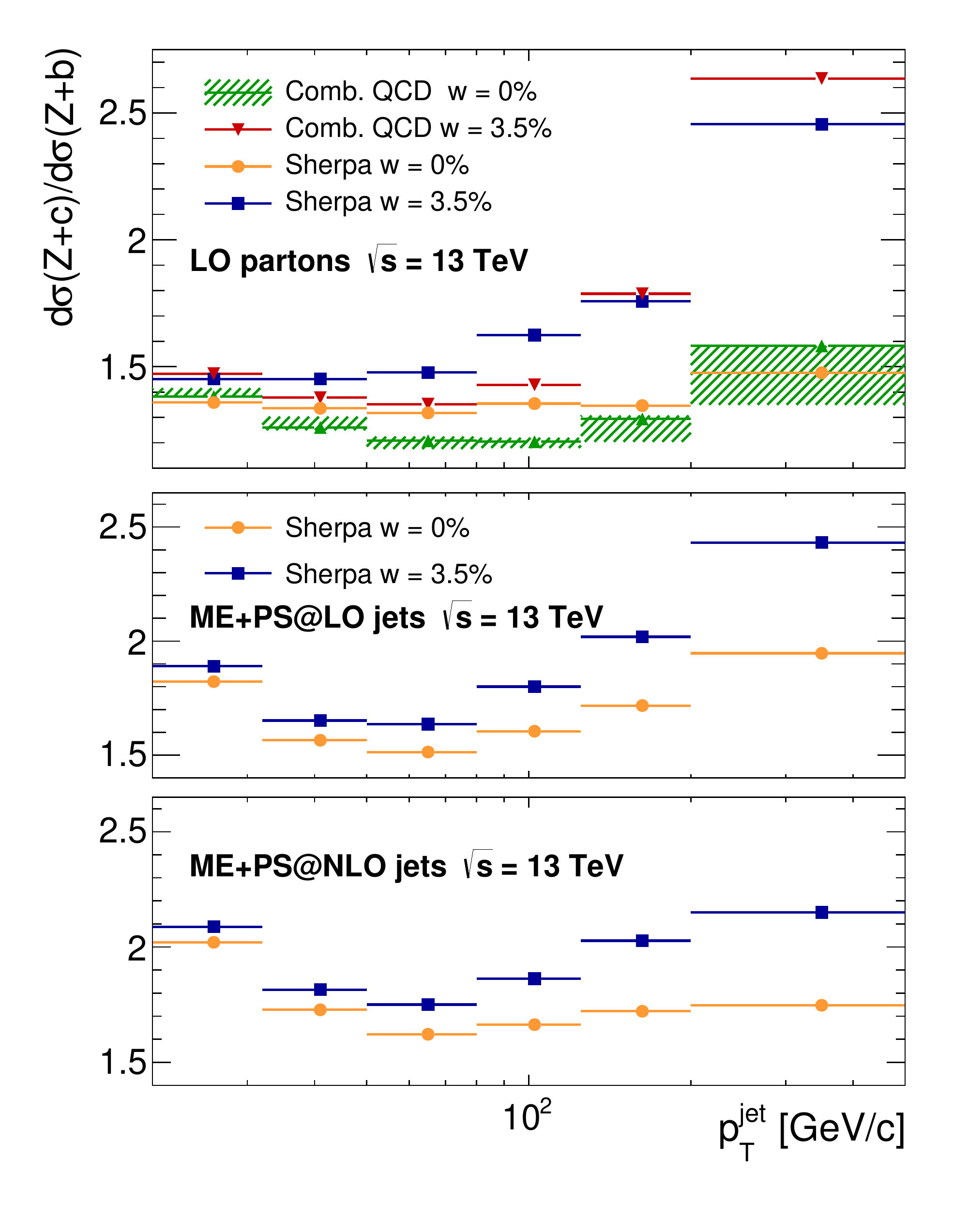}}\hfill
\caption{Predictions for the ratio of the production cross
sections of $Z+c$-jet and of $Z+b$-jet 
as a function of the $Z$ boson (left) and \HF jet (right)
transverse momenta in the forward rapidity region 
$1.5 < |y^{Z}| < 2.5$ at $\sqrt{s} = 13$~TeV. The predictions are made with
the \sherpa generator at the parton level
using the LO matrix element (top panels) and at
the particle level using the ME+PS@LO
(middle panels) and ME+PS@NLO (bottom panels) models. Predictions of the
combined QCD are also shown in the top panel. CTEQ66(c) PDF sets are used
with IC contribution values $w = 0$ and 3.5\%.}
\label{fig:explanation_IC_sup}
\end{center}
\end{figure}

Next, the effects of adding 
parton showers and NLO corrections to the parton 
level \sherpa LO predictions for differential cross section
ratios $\sigma{(Z + c)}/\sigma{(Z + b)}$ are illustrated in 
Fig.~\ref{fig:explanation_IC_sup}.
These ratios are calculated using CTEQ66(c) PDF sets with $w_{IC} = 0$\% and 
$3.5$\%. 
One can see that including parton showers does 
significantly decrease the excess in the spectrum caused by
the non-zero IC component, while adopting the
ME+PS@NLO instead of the ME+PS@LO approach makes little
difference. 
Thus, both \sherpa predictions made at a particle level give
the IC effect in the forward region at $200 < \pT < 500$~GeV
(irrespectively of whether \pT of the jet    
or of the $Z$ boson is considered) of the 
order of $10$ -- $20$\%, compared to the much larger 
effect predicted by the parton-level calculations (\sherpa at LO or
the combined QCD approach) to be at the level
of a factor of about 2. This observation is
in qualitative agreement with that 
made in~\cite{Hou:2017khm} when comparing the predictions for integral cross 
sections of $Z+c$-jet production from fixed order MCFM calculations and those 
from \sherpa within the ME+PS@LO approach.

\begin{figure}
\begin{center}
{\includegraphics[width=.5\textwidth]{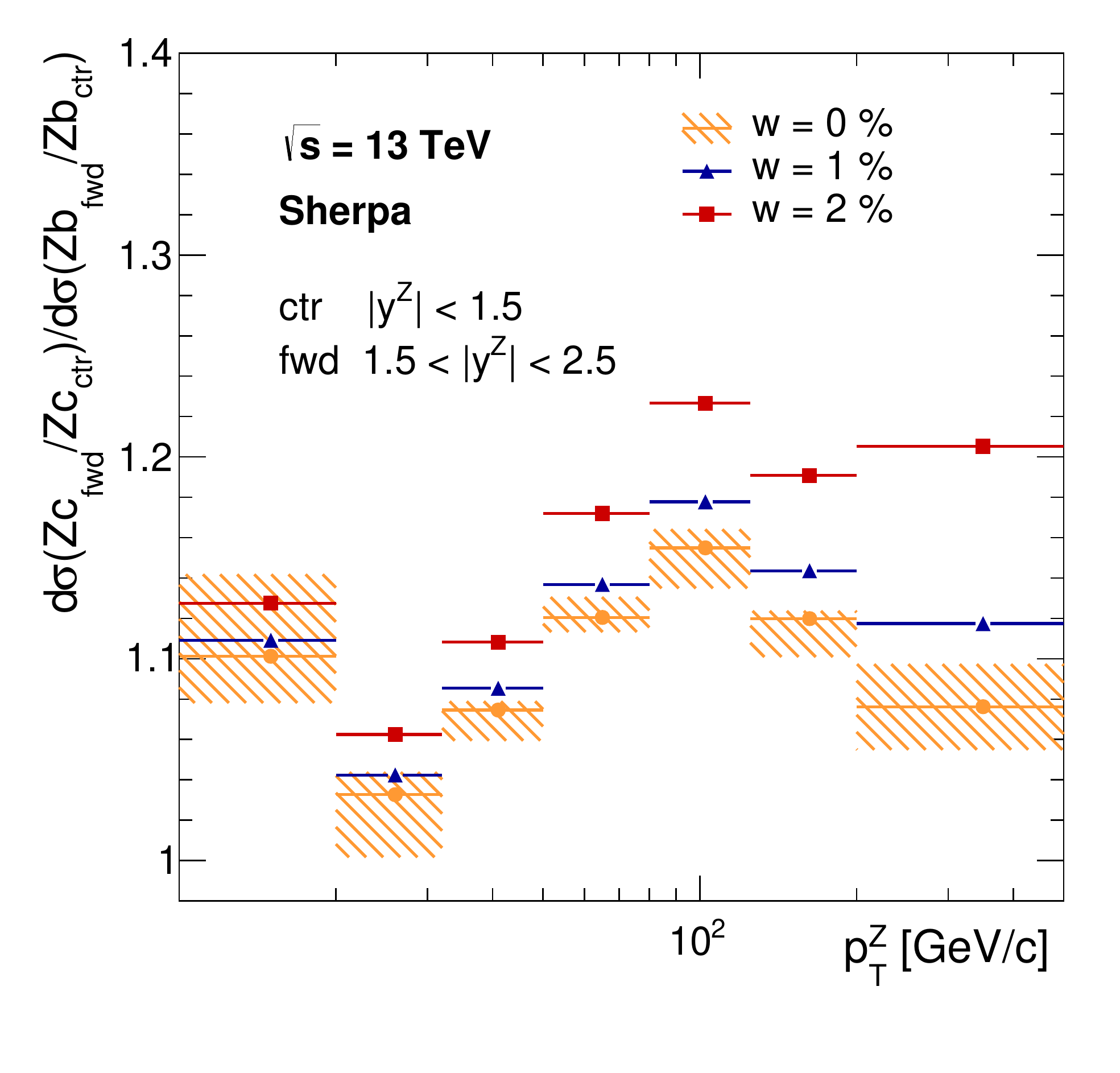}}\hfill
{\includegraphics[width=.5\textwidth]{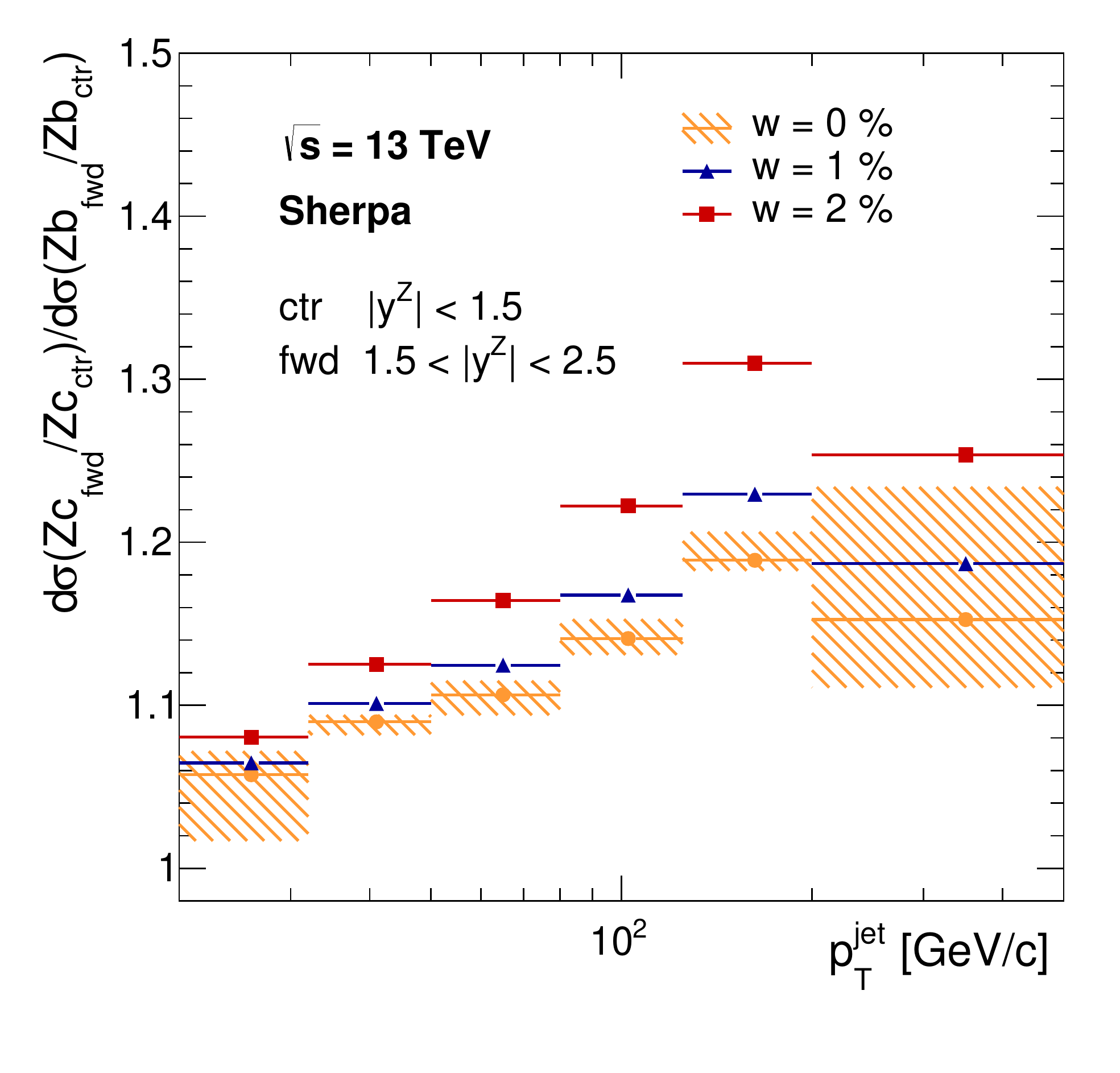}}\hfill
\caption{Predictions for the double ratio as a function of the
$Z$ boson (left) and jet (right) transverse momenta  at $\sqrt{s} = 13$~TeV. 
The double ratio is the ratio of the $Z+c$-jet production
cross section in the 
forward region $|y^{Z}| < 1.5$ to the cross section in the
central region $1.5 < |y^{Z}| < 2.5$, divided by the same ratio for $Z+b$-jet
production. 
The predictions are made with the 
\sherpa generator using CT14nnlo PDF with different IC contribution 
values $w$.
The uncertainty bands represent the uncertainties in the
QCD scale (shown only for $w=0\%$ predictions). 
}
\label{fig:double_ratio}
\end{center}
\end{figure}

Now we turn to the discussion of our theoretical uncertainties and 
uncertainties of the LHC measurements. These uncertainties
have been shown~\cite{Bednyakov:2017vck} to impose a strong
restriction on the precision of the IC probability estimation from the
experimental data. So new observables which may be less affected by such
uncertainties are of high interest. A new variable satisfying this criterion 
can be defined as follows. The ATLAS and CMS rapidity range
is divided into a central region $|y^{Z}|<1.5$ and
a forward region $1.5<|y^{Z}|<2.5$. Then, the 
ratio of the $Z+c$ production cross sections in
the forward region and in the central region 
is divided by the same ratio for $Z+b$ production. This
so-called double ratio 
$\sigma(Zc_{fwd}/Zc_{ctr})/\sigma(Zb_{fwd}/Zb_{ctr})$ is shown in 
Fig.~\ref{fig:double_ratio} as a function of the transverse 
momentum of the $Z$ boson $\pT^{Z}$ at the left and
of the leading jet $\pT^{\jet}$ at the right. 
One can see that in those ratios the IC effect
is already visible at the transverse momentum
$\pT \gtrsim 50$~GeV. 
This value is much less than if one studies the differential cross 
sections of $Z+c$ production.
Moreover, the uncertainties related to the QCD scale in
theoretical calculations are significantly suppressed in this
double ratio (see Fig.~\ref{fig:double_ratio}).
Therefore, the latter could be a more promising variable in the 
search for intrinsic  
charm at LHC as compared to other observables considered
previously.

Moreover, to obtain more reliable information on the probability 
of IC being present in the proton from future LHC data
at $\sqrt{s}= 13$~TeV one can perform a better estimation of
theoretical scale uncertainties and reduce systematic
uncertainties. This problem can be addressed by employing the
``principle of maximum conformality'' (PMC)~\cite{Brodsky:2013vpa} which sets
renormalization scales by shifting the $\beta$ terms in the pQCD series into
the running coupling. The PMC predictions are independent of
the choice of renormalization scheme -- a key
requirement of the renormalization group. 
However, up to now there is no direct
application of the PMC to the hard processes discussed in this paper.
One can expect forthcoming ATLAS and CMS 
experimental results on associated \ZHF production
to be sensitive to the effect of IC in a proton.

\section{Conclusion}
Associated production of the $Z$ boson and heavy flavor
jets in $pp$ collisions at LHC energies has
been considered applying the \sherpa Monte Carlo 
generator and the combined QCD factorization approach using PDF sets with
different intrinsic charm components. The combined QCD
approach employs both the \kT-factorization and the collinear QCD
factorization with each of them used in the kinematical conditions of its
reliability. The best description of the ATLAS and CMS data on the 
$Z + b$ and $Z + c$ production at $\sqrt{s} = 7$ and $8$~TeV was obtained
within the \sherpa 5FS ME+PS@NLO model. Effects arising from
parton showers and higher-order pQCD corrections have been
investigated. We found these effects to strongly suppress
the sensitivity of our predictions to the intrinsic charm content of a proton.
However, despite this suppression, one can expect forthcoming
ATLAS and CMS measurements of \ZHF production at $\sqrt{s} = 13$~TeV 
to be very important to search for the IC contribution in
the proton. We suggest to measure a new
observable, namely, the double ratio of cross 
sections $\sigma(Zc_{fwd}/Zc_{ctr})/\sigma(Zb_{fwd}/Zb_{ctr})$, which is extremely 
sensitive to the IC signal. This observable can be very promising
for precision estimation of the IC
probability, since it is less affected
by QCD scale uncertainties, as compared to
the observables considered previously.

\section*{Acknowledgments}
We thank V.A.~Bednyakov, S.J.~Brodsky, and F.~Sforza for extremely helpful
discussions and recommendations in the study of this topic. 
The authors are grateful to S.P.~Baranov, H.~Jung and P.M.~Nadolsky
for very useful discussions and comments.  
A.V.L. and M.A.M. are grateful to the DESY Directorate for 
support within the framework of the Moscow --- DESY project on Monte-Carlo 
implementation for HERA --- LHC.
M.A.M. was also supported by a grant of the
foundation for the advancement of theoretical physics and
mathematics “Basis” 17-14-455-1.

\bibliography{Paper_Z_plus_HF_2017}

\end{document}